\RequirePackage{fix-cm}

\documentclass[twocolumn,epjc3]{svjour3}  

\smartqed  

\RequirePackage{graphicx}

 \RequirePackage{mathptmx}      

\journalname{Eur. Phys. J. C}

\usepackage{xcolor}

\newcommand{\be}{\begin{equation}}
\newcommand{\ee}{\end{equation}}

\def\be{\begin{equation}}
\def\ee{\end{equation}}
\def\bea{\begin{eqnarray}}
\def\eea{\end{eqnarray}}

\usepackage{amssymb}

\begin{document}

\title{Spacetime mappings of the Brown-York quasilocal energy}


\author{Jeremy C\^ot\'e\thanksref{e1,addr1}
        \and
	Marianne Lapierre-L\'eonard\thanksref{e2,addr1}
	\and
        Valerio Faraoni\thanksref{e3,addr1} 
}

\thankstext{e1}{e-mail: jcote16@ubishops.ca}
\thankstext{e2}{e-mail: mlapierre12@ubishops.ca}
\thankstext{e3}{e-mail: vfaraoni@ubishops.ca}


\institute{Department of Physics \& Astronomy, Bishop's University, 
2600 College Street, Sherbrooke, Qu\'ebec, Canada J1M~1Z7 \label{addr1}
}

\date{Received: date / Accepted: date}

\maketitle

\begin{abstract}

In several areas of theoretical physics it is useful to know how a 
quasilocal energy transforms under conformal rescalings or generalized 
Kerr-Schild mappings. We derive the transformation properties of the 
Brown-York quasilocal energy in spherical symmetry and we contrast them 
with those of the Misner-Sharp-Hernandez energy.

\end{abstract}

\section{Introduction}
\label{sec:1}
\setcounter{equation}{0}

The mass of a non-isolated system in General Relativity 
(GR) has been the subject of intense study but there is no 
agreement as to what the mass-energy should be. Due to the 
equivalence principle, the energy of the gravitational 
field cannot be  localized and the mass-energy of a 
self-gravitating system includes also this energy. Unless 
the geometry reduces asymptotically to Minkowski (in which 
case the ADM energy \cite{ADM} is appropriate), one resorts 
to quasilocal energy definitions. 
  There are 
several quasilocal constructs in the literature, which 
differ from each other (see \cite{Szabados} for a recent 
review). Overall, quasilocal energy has been studied in the 
domain of formal relativity, but one ought to do 
better. First, the mass of a gravitating system is one of 
its most basic properties in astrophysics and a mass-energy 
definition is ultimately of no use if it cannot be employed 
in 
practical calculations (for example, in  
astrophysics and/or in cosmology). Second, various authors 
are already using, implicitly, the Hawking-Hayward 
quasilocal energy \cite{Hawking,Hayward} (usually in its 
Misner-Sharp-Hernandez 
form defined for  spherical symmetry \cite{MSH})
in black hole thermodynamics \cite{BHthermo}, in which this 
quasilocal energy plays the 
role of the internal energy of the system.

Black hole thermodynamics (especially the thermodynamics 
of time-dependent apparent horizons) is usually studied in 
the context 
of GR and most often in spherical symmetry, where the 
Misner-Sharp-Hernandez mass is adopted almost universally 
 \cite{BHthermo} (see, however, Ref.~\cite{VillalbaBargueno} for an 
analogous 
study using the Brown-York mass). The 
Misner-Sharp-Hernandez 
mass is also the quasilocal construct used in spherical 
fluid dynamics and in black hole collapse \cite{MSH} and is 
the Noether charge associated with the covariant conservation of the 
Kodama energy current \cite{Haywardspherical}. But 
there are 
several other definitions of quasilocal energy 
\cite{Szabados} and one wonders what changes the use of  
another quasilocal construct, for example the Brown-York 
energy, would bring. When a black hole is dynamical, it is 
difficult to calculate its temperature unambiguously 
and the recent 
literature on the thermodynamics of dynamical black holes 
focuses on this quantity. 
If the definition of internal energy is also uncertain, the 
problems accumulate. Quasilocal energies have been 
used also in the now rather broad field of thermodynamics 
of spacetime \cite{SptThermo}.

A full discussion of which quasilocal mass should be used, 
and why,   
requires more insight on quasilocal energies than is 
presently available. Here we consider a particular aspect, 
more related to tool-building than to core issues, 
which has been discussed recently in the literature. Since 
several analytic solutions of the Einstein field equations 
which describe dynamical black holes are generated by using 
the Schwarzschild (or another static black hole) solution 
as a seed and performing a conformal  or a 
Kerr-Schild transformation \cite{solutions}, the 
transformation properties 
of the Misner-Sharp-Hernandez mass under these spacetime 
mappings were discussed \cite{Enzo2013}. Later, 
relinquishing the simplifying assumption of spherical 
symmetry, the transformation properties of the 
Hawking-Hayward quasilocal energy \cite{Hawking,Hayward} 
(which reduces to the Misner-Sharp-Hernandez prescription 
\cite{MSH} in spherical symmetry \cite{Haywardspherical}) 
were also obtained \cite{HHconfo}. Generalizations of the 
Hawking-Hayward energy to scalar-tensor gravity have also 
been introduced (\cite{STquasilocal,HHFaycal}, see also 
\cite{otherSTquasilocal}, and \cite{Hideki} for the 
case of Lovelock gravity), following earlier 
generalizations of the Brown-York mass to these theories \cite{STBY}. A 
useful trick consists of remembering 
that these theories admit a representation in the 
Einstein conformal frame which is formally very similar to 
GR. If one chooses a different quasilocal energy, it 
becomes important 
to establish how 
this construct transforms under these spacetime 
mappings.

There are also other motivations for studying the 
transformation properties of quasilocal energies. As 
noted above, these 
quantities are defined rather formally and are not yet used 
in practical calculations in astrophysics and cosmology, 
with the exception of the recent works 
\cite{Nbody,turnaround,Lambdalensing}. There, the Hawking-Hayward 
quasilocal construction was employed in a new approach to 
cosmological problems  in which the expansion of the 
universe competes with the local dynamics of 
inhomogeneities, namely the Newtonian simulations of large 
scale structure formation in the early universe 
\cite{Nbody}, the 
turnaround radius in the present 
accelerated universe \cite{turnaround}, 
and lensing by the cosmological constant or by dark 
energy \cite{Lambdalensing}. To first order in the metric perturbations  
present in these problems, the Brown-York energy yields the same results 
as the Hawking-Hayward energy, provided that an appropriate gauge is 
chosen for the gauge-dependent Brown-York energy in the comparison 
\cite{new}.  Conformal 
transformations were used in these works as a mere calculational tool, 
not for any conceptual reason. This is one more 
reason to establish how the Brown-York mass behaves under 
spacetime mappings, if it was going to replace the 
Hawking-Hayward/Misner-Sharp-Hernandez construct.

In this work we restrict to spherical symmetry and, 
correspondingly, to a line 
element expressed in the gauge \cite{footnote1} 
\be
ds^2= -A(t,R)dt^2+B(t,R)dR^2 +R^2 d\Omega_{(2)}^2 \,,
\label{lineelement1}
\ee
where $R$ is the areal radius, a well-defined geometric 
invariant once spherical symmetry is assumed, and 
$d\Omega_{(2)}^2 \equiv d\theta^2 +\sin^2 \theta \, d\varphi^2$ is the 
line 
element on the unit 2-sphere. It is well known that, {\em in this gauge}, 
the Brown-York mass is given by \cite{BrownYork,BrownYorkLau,BlauRollier}
\be
M_{BY}=R\left( 1-\frac{1}{ \sqrt{B}} \right) \,. 
\label{BYmass}
\ee
The general definition of Brown-York mass 
is based on an integral of the extrinsic curvature of a 
3-surface in the real space minus the same quantity evaluated  
on the same Riemannian surface but with a Riemannian 3-space as a 
reference \cite{BrownYork}. It 
is clear from the definition that the Brown-York mass is gauge-dependent. 
By contrast, the Misner-Sharp-Hernandez mass $M_{MSH}$ is 
given by the 
{\em scalar} equation 
\be
1-\frac{2M_{MSH}(R)}{R}=\nabla^c R \nabla_c R
\ee
and is, therefore, gauge-independent, which is a significant 
practical advantage over the Brown-York mass.

Since the Brown-York mass is gauge-dependent, it 
makes sense to derive its transformation properties under 
conformal and Kerr-Schild spacetime mappings only when a 
certain gauge is preserved by the map. 
This is  what we do in the 
following sections. Since both the expression and 
the value of the Brown-York mass are very different in different gauges, 
it is meaningless to compare them in these different gauges.

\section{Conformal transformations}
\label{sec:2}

A conformal transformation of the metric is the 
point-dependent rescaling
\be
g_{ab} \rightarrow \tilde{g}_{ab}=\Omega^2 g_{ab} \,,
\ee
where the conformal 
factor $\Omega$ is a smooth positive function of 
the 
spacetime point. We require the conformal factor to respect 
the spherical symmetry, $\Omega=\Omega(t, R)$. 
Under such a mapping, the line 
element~(\ref{lineelement1}) becomes
\begin{eqnarray}
d\tilde{s}^2=\Omega^2 ds^2 =-\Omega^2 A 
dt^2+\Omega^2 B dR^2 
+\tilde{R}^2 
d\Omega_{(2)}^2 \,, \nonumber\\
\label{le:3}
\end{eqnarray}
where the ``new'' areal radius is $\tilde{R}=\Omega R$. The  
line element~(\ref{le:3}) is not in the 
form~(\ref{lineelement1}). To bring it to this form with 
tilded quantities, {\em i.e.}, 
\be
d\tilde{s}^2= -\tilde{A} \, d\tilde{t}^2+\tilde{B}d\tilde{R}^2 
+\tilde{R}^2 d\Omega_{(2)}^2 \,,
\label{le:4}
\ee
one has to introduce a new time coordinate. Begin by 
substituting the  differential 
\be 
dR= \frac{d\tilde{R}-\Omega_{,t} Rdt}{\Omega_{, R}R+\Omega }
\ee
in the line element~(\ref{lineelement1}), obtaining 
\begin{eqnarray}
d\tilde{s}^2 &=& -\Omega^2 \left[ A 
-\frac{ \Omega_{,t}^2R^2 B}{ 
\left( \Omega_{,R}R+\Omega \right)^2}\right] dt^2 
 +\frac{ \Omega^2B}{ \left( \Omega_{,R}R+\Omega \right)^2 }\, 
d\tilde{R}^2 \nonumber\\
&&\nonumber\\
&\, & 
-\, \frac{2\Omega^2 \Omega_{,t}BR}{ 
\left( \Omega_{,R}R+\Omega \right)^2}\, dtd\tilde{R} 
+\tilde{R}^2 d\Omega_{(2)}^2 \,. \label{11ter}
\end{eqnarray}
In order to bring this line element back to the 
gauge~(\ref{lineelement1}), one eliminates the term 
proportional to $dt \, 
d\tilde{R}$ by changing the 
time coordinate to $\tilde{t}(t, \tilde{R})$ defined by
\be \label{13}
d\tilde{t}=\frac{1}{F} \left( dt +\beta d\tilde{R} \right) \,,
\ee
where $\beta( t, R ) $ is a function to 
be determined in such a way that the $dtd\tilde{R}$ term 
disappears and $F (t, \tilde{R} )$ is a 
(non-unique)  
integrating factor satisfying
\be\label{eqforF}
\frac{\partial}{\partial \tilde{R}} \left( \frac{1}{F} \right) = 
\frac{\partial}{\partial t} \left( \frac{\beta}{F} \right) 
\ee 
to guarantee that $d\tilde{t}$ is an exact differential. 
Using $dt=Fd\tilde{t} -\beta d\tilde{R}$ in the line element gives
\begin{eqnarray}
d\tilde{s}^2 &=& -\Omega^2\left[ A -\frac{ 
\Omega_{,t}^2R^2 B}{ 
\left( \Omega_{,R}R+\Omega \right)^2}\right] F^2 d\tilde{t}^2  \nonumber\\
&&\nonumber\\
&\, &  + \left\{ -\beta^2 \left[ \Omega^2A -\frac{ 
\Omega_{,t}^2R^2\Omega^2 B}{ 
\left( \Omega_{,R}R+\Omega \right)^2}\right] 
+\frac{  \Omega^2 B}{ \left( \Omega_{,R}R+\Omega \right)^2 } 
\right.\nonumber\\
&&\nonumber\\
&\, & \left. 
+\frac{2 \beta \Omega_{,t}\Omega^2 BR}{ \left( \Omega_{,R}R+\Omega \right)^2}
\right\} d\tilde{R}^2 +2F \Omega^2 \cdot \nonumber\\
&&\nonumber\\
&\, &  \cdot \left\{ \beta \left[ A -\frac{ \Omega_{,t}^2R^2 
B}{ \left( \Omega_{,R}R+\Omega \right)^2}\right] - \frac{ 
\Omega_{,t} R B}{ \left( \Omega_{,R}R+\Omega \right)^2} 
\right\} d\tilde{t} \, d\tilde{R} \nonumber\\
&&\nonumber\\
&\, & +\tilde{R}^2 d\Omega_{{2}}^2 
\,. 
\end{eqnarray} 
By imposing that  
\be\label{16} 
\beta \left( t, R \right) = 
\frac{ \Omega_{,t} BR}{ \left[ A -\frac{ 
\Omega_{,t}^2R^2  B}{ \left( \Omega_{,R}R+\Omega 
\right)^2}\right] \left( \Omega_{,R}R+\Omega \right)^2} 
\ee 
the line element becomes 
\begin{eqnarray}
d\tilde{s}^2 &=&  
-\Omega^2 \left[ A -\frac{ \Omega_{,t}^2 BR^2}{ \left( 
\Omega_{,R}R+\Omega \right)^2}\right] F^2 d\tilde{t}^2 \nonumber\\
&&\nonumber\\
&\, &  + \frac{AB\Omega^2}{A\left( \Omega_{,R}R+\Omega
\right)^2 -\Omega_{,t}^2 BR^2}  d\tilde{R}^2 
+\tilde{R}^2 d\Omega_{{2}}^2 
\,.\label{17} 
\end{eqnarray}
Therefore, we have  
\be
\tilde{B}=\frac{AB\Omega^2}{A\left( \Omega_{,R}R+\Omega 
\right)^2 -\Omega_{,t}^2 BR^2} \,.
\ee
Using this expression, Eq.~(\ref{BYmass}) gives the 
Brown-York mass in the conformally rescaled world and in 
the chosen gauge
\be
\tilde{M}_{BY}=\Omega R \left[ 1- \frac{ \sqrt{ 
A\left( \Omega_{,R}R+\Omega \right)^2 -\Omega_{,t}^2 
BR^2}}{\sqrt{AB}\, \Omega } \right] \,. \label{newBY1}
\ee
In general, there is no simple expression of the ``new'' Brown-York 
mass in terms of the ``old'' one plus a simple correction, 
analogous to the one previously obtained for the 
Misner-Sharp-Hernandez 
mass \cite{Enzo2013}. One 
could, for example, rewrite Eq.~(\ref{newBY1}) as
\begin{eqnarray}
\tilde{M}_{BY} &=& \Omega M_{BY} \nonumber\\
&&\nonumber\\
& + & \frac{R}{ \sqrt{AB}} \left( 
\Omega \sqrt{A} -\sqrt{ A\left( \Omega_{,R}R+\Omega 
\right)^2 -\Omega_{,t}^2BR^2} \, \right) \label{cazzo2}
\end{eqnarray}
but this decomposition is arbitrary and not particularly 
enlightening anyway, even in the simplest situations in 
which the 
scale factor depends only on one of the variables 
$\left( t, R \right)$. For comparison, the transformation 
property of the Misner-Sharp-Hernandez mass under a 
conformal rescaling is \cite{Enzo2013}
\be \label{7}
\tilde{M}_{MSH}=\Omega M_{MSH} -\frac{R^3}{2\Omega}\, 
\nabla^a \Omega\nabla_a \Omega 
-R^2 \nabla^a \Omega  \nabla_a R \,.  \label{eq:16}
\ee
 The first term $\Omega M_{BY}$ in Eq.~(\ref{cazzo2}) can be 
interpreted by introducing  Newton's constant and remembering that, in a 
simple intepretation (dating back to Dicke) lengths 
and times scale with $\Omega$, while masses scale with $\Omega^{-1}$ 
\cite{Dicke}. However, the quasilocal energy is a complicated construct 
and cannot be expected to scale in a simple way under conformal 
transformations. As a consequence, the second term in the right hand side 
of Eq.~(\ref{cazzo2}) defies simple physical interpretation (the same can 
be said for the transformation property~(\ref{eq:16}) of the 
Misner-Sharp-Hernandez mass).

The effect of the transformation~(\ref{cazzo2}) on black hole 
thermodynamics is difficult to interpret. A Smarr relation was derived in 
Ref.~\cite{VillalbaBargueno} for vacuum, static, spherical black holes of 
the 
form~(\ref{lineelement1}):
\be
2TS= M_{BY}+2pA \,,
\ee
where $T$ and $S$ are the temperature and area of the event horizon, $S$ 
is the entropy, and 
\be
p= \frac{1}{8\pi} \left( \frac{A'_H}{2A_H \sqrt{B_H} } +\frac{1}{r_H 
\sqrt{B_H} } -\frac{1}{r_H} \right) \,, \label{Smarr}
\ee
while the subscript $H$ denotes quantities evaluated at the horizon. 
Assuming that, under a conformal transformation, $\tilde{T} \simeq 
T/\Omega $ in an adiabatic approximation (as argued in~\cite{myHawking}), 
$S=A/4$, and $\tilde{A}=\Omega^2 A$, 
Eq.~(\ref{Smarr}) would yield 
\begin{eqnarray}
2\tilde{T}\tilde{S} &=& \tilde{M}_{BY}-\frac{R}{\sqrt{AB}} \left( \Omega 
\sqrt{A} -\sqrt{ A \left( \Omega_{,R}R+\Omega \right)^2 -\Omega_{,t}BR^2 
} \,  \right) \nonumber\\
&&\nonumber\\
&\, &  +\frac{2p\tilde{A}}{\Omega}  
\end{eqnarray}
in the tilded world. Not much should be construed from this complicated 
relation between tilded quantities: a conformal transformation with 
$\Omega=\Omega(t,r)$ preserving the spherical symmetry has 
changed the situation in which Eq.~(\ref{Smarr}) was derived 
\cite{VillalbaBargueno}. 
Vacuum is no longer vacuum and, if $\Omega_{,t}\neq 0$, the black hole is 
not static, the event horizon is no longer present, and the notion of 
black hole is now defined by an apparent (instead of event) horizon, which 
is not null \cite{Universe}. The time dependence of the (apparent) horizon 
must be taken into account even in an adiabatic approximation 
\cite{myHawking}. Therefore, simple statements on the effect of 
the conformal transformation on thermodunamics cannot be made.

\section{Kerr-Schild transformations}
\label{sec:3}

A generalized Kerr-Schild transformation has the form
\be
g_{ab} \rightarrow \bar{g}_{ab}=g_{ab}+2\lambda \, l_al_b 
\,, \label{KerrSchild}
\ee
where $\lambda $ is a positive function and 
$l^a $ is a 
null and geodesic vector field of  $g_{ab}$, that is, 
\be
g_{ab}l^a l^b=0 \,, \;\;\;\;\;\;  
l^a \nabla_a l^b=0 \,.
\ee
It is easy to see that $l^a $ is null and geodesic also 
with respect to  $\bar{g}_{ab}$:
\be
\bar{g}_{ab} l^a l^b = g_{ab}l^a l^b +2\lambda (l_c l^c)^2 
 =0 \,, 
\ee
\be
\bar{g}_{ab} l^a \nabla^b l^c = 0 \,,
\ee
and that the inverse metric of $\bar{g}_{ab}$ is 
$ \bar{g}^{ab}= g^{ab}-2\lambda l^al^b $
since $  \bar{g}^{\mu\nu} \bar{g}_{\nu\alpha} =  
\delta^{\mu}_{\alpha}$ \cite{Enzo2013}. In order to 
respect the spherical symmetry of the 
geometry~(\ref{lineelement1}) we require that 
$\lambda=\lambda(t, R)$ and 
\be
l^{\mu} (t, R)=\left( l^0, l^1, 0,0 \right)
\ee
in this gauge. The generalized Kerr-Schild 
transformation~(\ref{KerrSchild}) gives 
\begin{eqnarray}
d\bar{s}^2 &=& ds^2 +2\lambda l_al_b dx^a dx^b \nonumber\\
&&\nonumber\\
&=& 
-\left[ A- 2\lambda (l_0)^2 \right] dt^2 
 +\left[ B+2\lambda 
(l_1)^2 \right] dR^2 +4\lambda l_0 l_1 dtdR \nonumber\\
&&\nonumber\\
& \, & +R^2  d\Omega_{(2)}^2  \,.
\end{eqnarray}
We now repeat the procedure 
of  Ref.~\cite{Enzo2013} in   
order to eliminate the cross-term in $dtdR$. To this end, 
it is necessary to introduce a new time coordinate $T$ 
defined by 
\be
dT=\frac{1}{F} \left( 
dt+\beta dR \right) \,,
\ee
where $\beta (t, R)$ is a function to be 
determined and $F(t, R)$ is an integrating factor. 
The substitution of $dt=FdT-\beta dR$ into the line element 
yields
\begin{eqnarray}
d\bar{s}^2 &=&  
-\left[ A-2\lambda (l_0)^2 \right] F^2 dT^2 \nonumber\\
&&\nonumber\\
&\, &  +\left\{ B+2\lambda (l_1)^2 -\beta^2 \left[ A-2\lambda (l_0)^2 
\right] - 4\lambda l_0 l_1 \beta \right\}  dR^2 \nonumber\\
&&\nonumber\\
&\, &  + 2F \left\{ 
\beta \left[ A-2\lambda (l_0)^2 \right] +2\lambda l_0 l_1 \right\} dTdR 
+  R^2 d\Omega_{(2)}^2  \,,\nonumber\\
&&
\end{eqnarray}
from which one deduces that the required form of the 
function $\beta$ is 
\be
\beta (t, R)= \frac{-2 \lambda l_0 l_1}{A-2\lambda (l_0)^2} 
\,.
\ee
With this choice, the metric is brought back to the 
gauge~(\ref{lineelement1}),
\begin{eqnarray}
d\bar{s}^2 &=& 
-\left[ A-2\lambda (l_0)^2 \right] F^2 dT^2  \\
&&\nonumber\\
&\, &  +
\left\{ B+ 2\lambda (l_1)^2 
+ \frac{ 4\lambda^2 (l_0)^2 (l_1)^2 }{ 
A-2\lambda (l_0)^2 } \right\}  dR^2 
+R^2 d\Omega_{(2)}^2  \,, \nonumber
\end{eqnarray}
where we note that
\be
\bar{B}= B+\frac{ 2\lambda A (l_1)^2}{A-2\lambda (l_0)^2} 
\ee
and there is residual gauge freedom due to the 
non-uniqueness of the integrating factor $F$. 
The Brown-York  mass of 
the barred spacetime is then given by the 
expression~(\ref{BYmass}) as 
\be \label{BY-KS}
\bar{M}_{BY}= R \left( 1- \frac{1}{ \sqrt{ B+ \frac{  
2\lambda A (l_1)^2}{A- 2\lambda (l_0)^2 } } } \right) \,.
\ee
Because of the normalization $l_c l^c=0$ of the null vector 
$l^a$, it is possible to rescale its components so  that, 
say, $ l_0=-1$. Then 
\be
l^{\mu}=\left( \frac{1}{A}, \frac{\pm 
1}{\sqrt{AB}}, 0,0 \right)
\ee
and Eq.~(\ref{BY-KS}) simplifies to
\begin{eqnarray}
\bar{M}_{BY} &=& R \left( 1- \frac{ \sqrt{ A- 
2\lambda }}{ \sqrt{AB}} \right) \nonumber\\
&&\nonumber\\
&=&  M_{BY}+\frac{R}{\sqrt{B}} \left( 1-\sqrt{ 
1-\frac{2\lambda}{A} }\right)\,.  \label{cazzoquadro}
\end{eqnarray}
For comparison, the transformation
property of the Misner-Sharp-Hernandez mass under a
generalized Kerr-Schild map is \cite{Enzo2013}
\be \label{MSHmassKS}
\bar{M}_{MSH}= M_{MSH} +\frac{ \lambda R}{AB} \,.
\ee
Again, the action of the Kerr-Schild transformation arising from a 
nonvanishing $\lambda$ cannot be given a simple interpretation due to the 
fact that quasilocal energies are rather complicated constructs.

\section{Examples}
\label{sec:3bis}

Here we present examples illustrating the transformation properties of 
the Brown-York mass.

\subsection{Conformal transformation}

Consider the Minkowski line element in polar coordinates
\be
ds^2 =-d\eta^2 +dr^2+r^2 d\Omega_{(2)}^2 \,, \label{Minkowski}
\ee
where $r=R$ is trivially the areal radius. The spatially flat 
Friedmann-Lema\^itre-Robertson-Walke (FLRW) line element 
\be
d\tilde{s}^2=a^2(\eta) \left( -d\eta^2 +dr^2 +r^2 d\Omega_{(2)}^2 \right) 
\, \label{FLRW}
\ee
where $\eta$ is the conformal time, is manifestly conformally 
flat. The 
conformal transformation $d\tilde{s}^2 =\Omega^2 ds^2$ relating 
(\ref{Minkowski})  and (\ref{FLRW}) has conformal factor 
$\Omega=a(\eta)$, the scale factor of 
the universe. The FLRW areal radius is $\tilde{R}=a(\eta)r $ and the 
Hubble parameters in comoving time $t$ (given by $dt=ad\eta$) and 
conformal time $\eta$ are, respectively,  $H\equiv \dot{a}/a$ and ${\cal 
H}=a_{\eta}/a=aH 
$ (an overdot 
denoting differentiation with respect to $t$). 

To write the FLRW line element~(\ref{FLRW}) in Schwarzschild-like 
coordinates, we introduce the new time $T$ by
\be
dT=\frac{1}{F} \left( dt+\beta d\tilde{R} \right) \,,
\ee
which transforms~(\ref{FLRW}) to 
\begin{eqnarray}
d\tilde{s}^2 &=& -\left( 1-H^2 \tilde{R}^2 \right) F^2 dT^2  
\nonumber\\
&&\nonumber\\
&\, & -2F \left[ 
-\left( 1-H^2 \tilde{R}^2 \right)\beta +H\tilde{R} \right] dT d\tilde{R} 
\nonumber\\
&&\nonumber\\
&\, &  +
\left[ -\left( 1-H^2 \tilde{R}^2 \right)\beta^2 +2  H\tilde{R} \beta 
+1\right] d\tilde{R}^2 + \tilde{R}^2 d\Omega_{(2)}^2 \,. \nonumber\\
&& 
\end{eqnarray}
The choice 
\be
\beta(t, \tilde{R}) = \frac{H\tilde{R}}{1-H^2\tilde{R}^2} 
\ee
reduces the line element to the Schwarzschild-like gauge~(\ref{le:4}) with 
$\tilde{A}= \left(  1-H^2 \tilde{R}^2 \right)F^2$, $ \tilde{B}= \left( 
1-H^2 \tilde{R}^2  \right)^{-1}$. As  a consequence, the 
expression~(\ref{cazzo2}) of the Brown-York mass in 
spherical symmetry yields 
\begin{eqnarray}
\tilde{M}_{BY}^{(FLRW)} &=& \tilde{R} \left( 1-\sqrt{ 1-H^2 \tilde{R}^2} 
\right) = \tilde{R} \left( 1-\sqrt{1-\frac{8\pi}{3}\, \rho \tilde{R}^2} 
\right)\nonumber\\
&&\nonumber\\
& = & 
\tilde{R} \left( 1-\sqrt{1-\frac{2M_{MSH}^{(FLRW)} }{\tilde{R}} } \right) 
\,,\label{lolli}
\end{eqnarray}
where $M_{MSH}^{(FLRW)} =H^2\tilde{R}^2/2 = \frac{4\pi}{3} \, \rho 
\tilde{R}^3$ is the Misner-Sharp-Hernandez mass of FLRW space and we used 
the Friedmann equation $H^2=8\pi  \rho/3$, with $\rho$ the 
energy density of the cosmic fluid. The relation~(\ref{lolli}) between the 
two quasilocal masses in this example mirrors that holding in the 
Schwarzschild geometry
\be
ds^2=-\left( 1-\frac{2m}{r} \right) dt^2 + \frac{dr^2}{1-2m/r} +r^2 
d\Omega_{(2)}^2
\ee
with (constant) mass $m$. The Misner-Sharp-Hernandez mass contained in a 
sphere of radius $r$ is $M_{MSH}^{(Schw)}= m$ for any value of $r>2m$ and  
the Brown-York mass is
\be
M_{BY}^{(Schw)}= r\left( 1-\sqrt{1-\frac{2m}{r}} \right) \,,
\ee
and it asymptotes to $m$ as $r\rightarrow +\infty$. However, on the 
Schwarzschild event 
horizon $r=2m$, it is $M_{BY}^{(Schw)}=2 M_{MSH}^{(Schw)}$.

\subsection{Kerr-Schild transformation}

As an example of Kerr-Schild transformation, consider the map between 
the Minkowski geometry~(\ref{Minkowski}) and the Reissner-Nordstr\"{o}m 
one
\begin{eqnarray}
d\tilde{s}^2 &=& -\left( 1-\frac{2m}{r}+\frac{q^2}{r^2} \right) dt^2 
+\frac{dr^2}{1-\frac{2m}{r}+\frac{q^2}{r^2} } +r^2 d\Omega_{(2)}^2 
\,,\nonumber\\
&& 
\end{eqnarray}
which corresponds to
\be
\lambda(t,r)= \frac{m}{r}-\frac{q^2}{r^2} \,, \;\;\;\;\;\;\;\;\; 
l^{\mu}=\left( 1, -1, 0,0 \right) 
\ee
and to the time redefinition $t\rightarrow T$ with 
\be
dT=  dt -\, \frac{ \left( \frac{2m}{r}+ \frac{q^2}{r^2} \right)}{
1-\frac{2m}{r} +\frac{q^2}{r^2}  } \,  dr \,.
\ee 
Minkowski space has 
vanishing Brown-York mass and the 
transformation property (\ref{cazzoquadro}) of the Brown-York mass under 
Kerr-Schild transformations yields 
\be
\bar{M}_{BY}= r\left(  1- \sqrt{ 1- \frac{2m}{r}+\frac{q^2}{r^2} } \right) 
\,,
\ee
which coincides with the well known  expression calculated directly using 
Eq.~(\ref{BYmass}).

\section{Conclusions}
\label{sec:4}

There are several reasons to derive the transformation 
properties of a quasilocal energy under conformal or 
(generalized) Kerr-Schild transformations. This procedure 
is part of the tool-building process useful in various 
areas of theoretical gravity (black hole thermodynamics, 
analytical solutions of GR describing dynamical black 
holes, spacetime thermodynamics, {\em etc.}). The relativity 
community 
seems to have concentrated on the 
Hawking-Hayward/Misner-Sharp-Hernandez quasilocal energy  (see, however, 
Refs.~\cite{BlauRollier,YuCaldwell,VillalbaBargueno}) but the Brown-York 
energy is also interesting in principle because it is based 
on the Hamilton-Jacobi formulation of GR. 
However, contrary to the Misner-Sharp-Hernandez mass, the 
Brown-York constructs suffers from a daunting 
gauge-dependence even in spherical symmetry. For 
this reason, the comparison of the ``new'' Brown-York 
energy after a spacetime mapping with the ``old'' one is 
meaningful only after restoring the gauge which is altered 
by the spacetime mapping. Having done this and having 
obtained the ``new'' Brown-York mass in terms of the 
``old'' one and of the geometry, the result cannot be 
encapsulated in a simple formula analogous to 
Eqs.~(\ref{7}) or (\ref{MSHmassKS}) obtained for the 
Misner-Sharp-Hernandez mass in \cite{Enzo2013}. From a 
pragmatic point of view, the Misner-Sharp-Hernandez 
construct looks definitely more attractive than the 
Brown-York one.

\begin{acknowledgements}
We are grateful to a referee for constructing comments.  
This work is supported, in part, by the Natural Sciences and Engineering 
Research Council of Canada (Grant No. 2016-03803 to V.F.) and by Bishop's 
University.

\end{acknowledgements}

\end{document}